\documentclass[12pt]{iopart}
\usepackage{graphicx}
\usepackage[small,bf]{caption}
\begin{document}
\title{Security and entanglement in differential-phase-shift quantum key distribution}
\author{Adriana Marais $^1$, Thomas Konrad $^1$ and Francesco Petruccione $^2$}
\address{$^1$\ Quantum Research Group, School of Physics, University of KwaZulu-Natal, Durban 4001, South Africa}
\address{$^2$\ Quantum Research Group, School of Physics and National Institute for Theoretical Physics, University of KwaZulu-Natal, Durban 4001, South Africa}
\eads{adrianamarais@gmail.com}
\newcommand{\plus}{\ensuremath{\ +\ }}
\newcommand{\minus}{\ensuremath{\ - }}
\newcommand{\eq}{\ensuremath{\ =\ }}
\newcommand{\bone}{\ensuremath{\stackrel{BS1}{\rightarrow}}}
\newcommand{\btwo}{\ensuremath{\stackrel{BS2}{\rightarrow}}}
\newcommand{\intf}{\ensuremath{\stackrel{I}{\rightarrow}}}
\begin{abstract}
The differential-phase-shift quantum key distribution protocol is formalised as a prepare-and-measure scheme and translated into an equivalent entanglement-based protocol. A necessary condition for security is that Bob's measurement can detect the entanglement of the distributed state in the entanglement-based translation, which implies that his measurement is described by non-commuting POVM elements. This condition is shown to be met.  
\end{abstract}
\noindent{\it Keywords\/}: quantum key distribution, differential-phase-shift protocol, security, entanglement
\section{Introduction}
Quantum key distribution (QKD) is a means of distributing a secure key between two parties, traditionally Alice and Bob, who wish to communicate privately \cite{gisin02,scarani08}. The security of the resulting shared key is independent of a potential eavesdropper's computational and technological power, if used as a one-time pad \cite{vernam26}.\\ 
\\
In `prepare-and-measure (P$\&$M)' terminology, a QKD protocol involves Alice preparing a set of quantum states into which a sequence of symbols has been encoded. These states are then sent to Bob via a quantum channel, who performs a measurement that serves to decode the signal and estimate the noise introduced by the channel, which is conservatively attributed to eavesdropping \cite{scarani08}. A general QKD protocol can be described equivalently as an entanglement-based (EB) scheme \cite{bbm92}, involving the preparation of a bipartite entangled state. When Alice performs a measurement on her subsystem, she effectively prepares the same set of quantum states described in the P$\&$M picture, which is then measured by Bob with the same measurement.\\
\\
In both the P$\&$M and in the equivalent EB scheme, a public classical channel, of which Alice and Bob are the authenticated users, is subsequently employed to filter a secure key out of the initial sequence of symbols. The classical data produced in a secure QKD protocol must imply non-classical correlations \cite{acin05} between the systems held by Alice and Bob in the EB translation. Therefore, a necessary condition for the security of a QKD protocol is that the measurements performed by Alice and Bob in the EB translation must detect entanglement in the effectively distributed state \cite{curty04}, which in turn implies that Bob's measurement must consist of non-commuting POVM elements (see Appendix).\\
\\ 
QKD protocols can be divided into three classes: discrete-variable (DV), continuous-variable (CV) and distributed-phase-reference protocols \cite{scarani08}. Differential-phase-shift (DPS)QKD is an example of a distributed-phase-reference protocol.\\
\\
The form of DPSQKD discussed here was proposed in 2003 by Inoue et al.\ \cite{inoue03} as a scheme offering a higher key creation efficiency than conventional fibre-based BB84. In 2006 Waks et al.\ \cite{waks06} derived a proof of the security of DPSQKD under the assumption that Eve is restricted to individual attacks. They showed that individual attacks are more powerful than certain so-called sequential attacks, thus ensuring security against this form of attack also. In the same year Diamanti et al.\ \cite{diamanti06} reported an implementation of DPSQKD secure against individual attacks over 100km. In 2007 Tsurumaru \cite{tsurumaru07} introduced an improved version of the aforementioned sequential attack that decreases the distance over which DPSQKD is secure to less than 95km, thus rendering the above implementation insecure. In 2009 Ma et al.\ \cite{ma09} reported an experimental realisation of DPSQKD using superconducting single-photon detectors, with a quantum bit error rate of less than 4$\%$. Later in 2009, a proof of the unconditional security of a protocol related to DPSQKD, using single photons instead of coherent pulses, was published \cite{wen09}. However, this proof does not imply the unconditional security of the original DPSQKD protocol.\\
\\
There are a number of practical advantages to DPSQKD, namely: its suitability for fibre transmissions; use of readily available telecommunication tools; no requirement for a single photon source (the generated states are assumed to be easily produced coherent states) and thus high communication efficiency. However, bounds for the unconditional security of DPSQKD, and other examples of distributed-phase-reference protocols like the coherent-one-way (COW) protocol \cite{gisin04,stucki05}, have not yet been found.\\ 
\\
A number of techniques have been used to show the unconditional security of DV protocols \cite{mayers96,lo99,shor00,kraus05}, and security proofs for CV protocols are developing to a similar level \cite{navas05}. For DV and CV protocols, the notion of virtual entanglement plays an essential role in security proofs based on entanglement distillation which are applied to the EB version. The same security proofs then also directly apply to the equivalent P$\&$M scheme. Note that the EB version is not necessarily implemented (hence the word `virtual'), but serves as a theoretical tool owing to its equivalence to the P$\&$M scheme.\\
\\
An EB translation of a P$\&$M QKD protocol where the precondition for security on Bob's measurement is satisfied, is a necessary first step towards a potential unconditional security proof for the protocol based on entanglement distillation. The purpose of this paper is to propose and formalise an EB translation of DPSQKD, as well as to show that Bob's measurement contains non-commuting POVM elements. This is a necessary condition for the unconditional security of the EB translation.\\
\\ 
This paper is organised as follows: In section 2 of this article, DPSQKD is described as a P$\&$M scheme. In section 3 the P$\&$M description of DPSQKD is translated into an equivalent EB scheme. In section 4 a necessary requirement for security- that Bob's measurement is described by non-commuting POVM elements- is shown to be fulfilled.  
\section{DPSQKD as a P$\&$M Scheme}
A general P$\&$M protocol specifies a quantum state $|\Psi(S)\rangle$ that encodes a sequence of $N$ symbols $S=\{s_{1}, ...,s_{N}\}$ prepared by Alice. In CV and DV protocols, $|\Psi(S)\rangle$ can be written in tensor product form where there is a one-to-one correspondence between each symbol $s_{i}$ and each state $|\psi(s_{i})\rangle$ that encodes that symbol:
\begin{equation}
|\Psi(S)\rangle\eq\bigotimes^N_{i=1}|\psi(s_{i})\rangle.
\end{equation}
The $s_{i}$'s are independent, and the state sent in each time interval $i$ can therefore be considered independently as the state $|\psi(s_{i})\rangle$. Note that the states $|\psi(s_{i})\rangle$ must be non-orthogonal since a set of orthogonal states can be perfectly copied by a potential eavesdropper.\\
\\
In DPSQKD Alice prepares a sequence of $N+1$ symbols $S'=\{s'_{0}, ..., s'_{N}\},\ s'_{i}\in\{0,1\}$, according to which she modulates the phase of each of $N+1$ attenuated coherent pulses by $\{0,\pi\}$. The pulses are separated by time $\Delta t$. After modulation the phase of the $i^{\mathrm{th}}$ pulse is given by $\phi_{i}=s'_{i}\pi$. From the bit string $S'$ Alice calculates the potential key bit string $S$ via the relation $s_{i}=s'_{i-1}+s'_{i}$, where addition is modulo 2. Let the time intervals in which Alice sends the pulses be denoted by $i=0, ..., N$. The quantum state $|\Psi(S')\rangle_{DPS}$ that encodes the sequence $S'$ can then be written as a tensor product
\begin{equation}
|\Psi(S')\rangle_{DPS}=\bigotimes^{N}_{i=0}|(-1)^{s'_{i}}\alpha\rangle.
\label{statedps}
\end{equation}
The requirement of non-orthogonality of the states $|\psi(s'_{i})\rangle$ (compare with $|\psi(s_{i})\rangle$ from Eq. (1)) is met, since the coherent pulses have an average photon number $|\alpha|^2$ of less than one.\\
\\
The state sent in each time interval $i$ can be considered independently if written (as above) as $|(-1)^{s'_{i}}\alpha\rangle$. Here there is a one-to-one correspondence between each symbol $s'_{i}$ and each state $|\psi(s'_{i})\rangle$ encoding that symbol, but since the consecutive elements of $S$ are not independent there is no such correspondence between potential key bits and prepared states, as in the general case. And this is the reason existing methods of proving unconditional security cannot be applied to the DPSQKD protocol, since they rely on the mutual independence of all potential key bits. However, the form of Eq. (2) allows a formulation of P$\&$M DPSQKD as an equivalent EB scheme.\\
\begin{figure} 
\begin{center}
\includegraphics[scale=0.4]{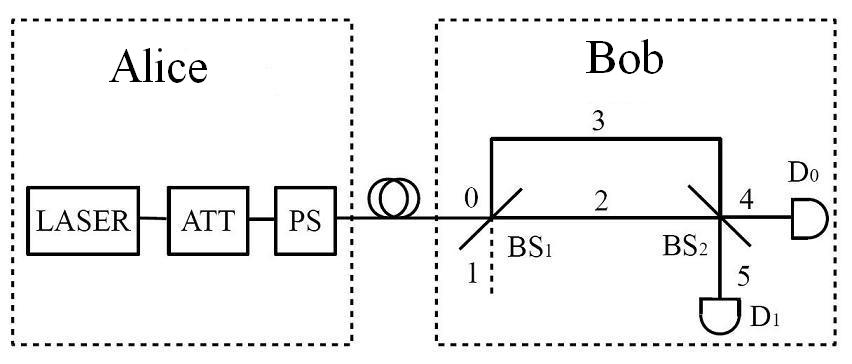} 
\end{center}
\caption[]{QKD system for the implementation of the DPSQKD protocol. LASER: coherent light source, ATT: attenuator, PS: phase shifter, BS: symmetric beamsplitter, D: detector. The $i^{\mathrm{th}}$ incoming pulse is split at BS$_1$. The part which propagates on path 3 arrives at BS$_2$ simultaneously with the part of the $(i+1)^{\mathrm{th}}$ pulse coming from path 2.
\label{figure1}}
\end{figure}
\\
Subsequent to Alice's preparation and sending of the state $|\Psi(S)\rangle$ to Bob in a general P$\&$M scheme, Bob performs a measurement which decodes the states. Measurement results contribute to the key, but are also used to estimate the loss of quantum coherence in the sent state. \\
\\
In DPSQKD Bob's measurement is initiated at a time $\Delta t$ after the first pulse has entered his interferometer, i.e., in time interval $i=1$, and there is the possibility of a detection event that can contribute to the key in this and subsequent time intervals up to $i=N$. The transformation of $|\Psi(S)\rangle_{DPS}$ in Bob's interferometer forms part of his measurement (see Fig. 1) and is described by the following transformations:\\
\\
The incoming pulses enter symmetric beam splitter 1 (BS$_1$) in the path labelled 0 and are described by field operators $\hat{a}^{\dagger}_{0}$, while path 1 contains vacuum. The beam splitter transformation for $\hat{a}^{\dagger}_{0}$ in terms of operators $\hat{a}^{\dagger}_{2}$ and $\hat{a}^{\dagger}_{3}$ for the output paths labelled 2 and 3 is
\begin{eqnarray}
\hat{a}^{\dagger}_{0}&\bone&\frac{1}{\sqrt{2}}\hat{a}^{\dagger}_{2}+\mbox{e}^{-i\phi_{1}}\frac{1}{\sqrt{2}}\hat{a}^{\dagger}_{3}.
\end{eqnarray}
Transformations for symmetric beam splitter 2 (BS$_2$) in terms of operators $\hat{a}^{\dagger}_{4}$ and $\hat{a}^{\dagger}_{5}$ for the output paths labelled 4 and 5 are given by
\begin{eqnarray}
\hat{a}^{\dagger}_{2}&\btwo&\frac{1}{\sqrt{2}}\hat{a}^{\dagger}_{4}-\mbox{e}^{i\phi_{2}}\frac{1}{\sqrt{2}}\hat{a}^{\dagger}_{5},\\
\hat{a}^{\dagger}_{3}&\btwo&\mbox{e}^{-i\phi_{2}}\frac{1}{\sqrt{2}}\hat{a}^{\dagger}_{4}+\frac{1}{\sqrt{2}}\hat{a}^{\dagger}_{5}.
\end{eqnarray}
The total action of the interferometer is described in terms of the creation operators for each time interval $i$ for the signals entering Bob's interferometer in path 0, $\hat{a}_{0}^{\dagger i}$, and the two outgoing paths, $\hat{a}_{4}^{\dagger i}$ and $\hat{a}_{5}^{\dagger i}$: 
\begin{eqnarray}
\hat{a}_{0}^{\dagger i}&\intf&\frac{1}{2}(\hat{a}_{4}^{\dagger i}-\mbox{e}^{i\phi_{2}}\hat{a}_{5}^{\dagger i}+\mbox{e}^{i(\phi_{\Delta t}-\phi_{1})}(\mbox{e}^{-i\phi_{2}}\hat{a}_{4}^{\dagger (i+1)}+\hat{a}_{5}^{\dagger (i+1)})),\nonumber\\
&=&\frac{1}{2}(\hat{a}_{4}^{\dagger i}-\mbox{e}^{i\phi_{2}}\hat{a}_{5}^{\dagger i}+\hat{a}_{4}^{\dagger (i+1)}+\mbox{e}^{i\phi_{2}}\hat{a}_{5}^{\dagger (i+1)}),
\end{eqnarray}
where the beam splitters are chosen such that their relative phase shifts compensate the phase shift associated with the time delay $\Delta t$ in the upper arm of the interferometer i.e., $\phi_1+\phi_2=\phi_{\Delta t}$. This time delay must be equal to the time separation between incoming pulses in path 0, so that components of consecutive pulses can interfere at BS$_2$.\\ 
\\
If Alice sends the state $|\Psi(S')\rangle_{DPS}$, the state $|\Psi '(S')\rangle_{DPS}$ entering Bob's detectors in time intervals $i$ in paths 4 and 5 after transformation in the interferometer, is expanded as
\begin{eqnarray}
|\Psi '(S')\rangle_{DPS}
&=&\bigotimes^{N}_{i=1}|\frac{1}{2}
\alpha(\mbox{e}^{i\phi_{i}}+\mbox{e}^{i\phi_{i-1}})\rangle_{4}^{i}|\frac{1}{2}\alpha\mbox{e}^{i\phi_{2}}(\mbox{e}^{i\phi_{i}}-\mbox{e}^{i\phi_{i-1}})\rangle_{5}^{i}\nonumber\\
&=&\bigotimes^{N}_{i=1}|\frac{1}{2}\alpha
((-1)^{s'_{i}}+(-1)^{s'_{i-1}})\rangle_{4}^{i}
|\frac{1}{2}\alpha\mbox{e}^{i\phi_{2}}((-1)^{s'_{i}}-(-1)^{s'_{i-1}})\rangle_{5}^{i},\nonumber\\
\end{eqnarray}
recalling that $\phi_{i}=s'_{i}\pi$.\\
\\
Bob uses detectors $D_0$ and $D_1$ that discern vacuum from one or more photons (so-called bucket detectors), in paths 4 and 5, respectively. A detector `click' will occur when one or more photons are detected. For an incoming coherent state $|\beta\rangle_4(|\beta\rangle_5)$, detector $D_0(D_1)$ will click with probability $1-\mbox{e}^{-|\beta|^{2}}$. These probabilities depend on the phase modulation performed by Alice, which is determined by the bit string $S'$. For $s_i'+s_{i-1}'=s_i=0(1)$, the probability of $D_1(D_0)$ clicking is zero, and hence a click in $D_0(D_1)$ corresponds to the potential key bit $s_i=0(1)$. This situation is summarised in Table 1.\\
\\   
\begin{table}
\begin{center}
\begin{tabular}{cccccc}
\hline
$s'_{i-1}$&$s'_{i}$&$s_{i}$&$|\psi\rangle^i$&$p(D_{0})$&$p(D_{1})$\\
\hline
0&0&0&$|\alpha\rangle_{4}|0\rangle_{5}$&$1-\mbox{e}^{-|\alpha|^{2}}$&0\\
1&1&0&$|-\alpha\rangle_{4}|0\rangle_{5}$&$1-\mbox{e}^{-|\alpha|^{2}}$&0\\
0&1&1&$|0\rangle_{4}|\alpha\mbox{e}^{\phi_{2}}\rangle_{5}$&0&$1-\mbox{e}^{-|\alpha|^{2}}$\\
1&0&1&$|0\rangle_{4}|-\alpha\mbox{e}^{\phi_{2}}\rangle_{5}$&0&$1-\mbox{e}^{-|\alpha|^{2}}$\\
\hline
\end{tabular}
\caption{DPSQKD detection and key extraction table: A coherent state $|\beta\rangle_4(|\beta\rangle_5)$ results in detector $D_0(D_1)$ clicking with probability $1-\mbox{e}^{-|\beta|^{2}}$. These probabilities depend on the phase modulation performed by Alice, which is determined by the bit string $S'$. For $s_i'+s_{i-1}'=s_i=0(1)$, the probability of $D_1(D_0)$ clicking is zero, and hence a click in $D_0(D_1)$ corresponds to the potential key bit $s_i=0(1)$.}
\end{center}
\end{table}
Note, the probability of a detector firing is independent of the phase, and therefore Bob cannot distinguish the two states $|\alpha\rangle^i$ and $|-\alpha\rangle^i$ that correspond to one $s_{i}$.\\
\\
Bob then utilises the authenticated classical channel to communicate to Alice the time intervals $i^*$ in which he recorded a detection event in one of his detectors (which is not in every time interval since the average photon number $|\alpha|^{2}$ per pulse is less than one). In the error-free case, this process serves to filter a secure key $S^*$ out of the initial sequence $S$, since for each $i^*$ Alice and Bob can add an identical bit, $s_{i^*}$, to the secure filtered key.
\section{DPSQKD as an EB Scheme}
The existence of equivalent EB translations for P$\&$M QKD schemes was first shown by Bennett et al.\ \cite{bbm92}. In the EB translation of a general P$\&$M protocol, the bipartite entangled state
\begin{equation}
|\Phi\rangle_{AB} \eq \frac{1}{\sqrt{D}}\sum_{S}|S\rangle_{A}\otimes|\Psi(S)\rangle_{B}
\end{equation}
is prepared, where $D$ is the number of possible $S$ sequences and the states $|S\rangle_{A}$ form an orthogonal basis for the $D$-dimensional space. By measuring in this basis, Alice learns one sequence $S$ and the corresponding $|\Psi(S)\rangle$ is effectively sent to Bob, who performs an identical measurement to that performed in the P$\&$M scheme.\\ 
\\
Since the $s_{i}$'s are independent (and let them be of an alphabet of size $d$), $|\Phi\rangle_{AB}$ can also be written as
\begin{equation}
|\Phi\rangle_{AB}=\bigotimes^N_{i=0}\left(\frac{1}{\sqrt{d}}\sum_{s_{i}}|s_{i}\rangle_{A}\otimes|\psi(s_{i})\rangle_{B}\right)
\end{equation}
where the states $|s_{i}\rangle$ form an orthogonal basis for a $d$-dimensional space.\\
\\
In the EB translation of DPSQKD, the bipartite entangled state
\begin{eqnarray}
|\Phi_{DPS}\rangle_{AB}&=& \frac{1}{\sqrt{2^{N+1}}}\sum_{S'}|S'\rangle_{A}\otimes(|\Psi(S')\rangle_{DPS})_{B}\nonumber\\
&=&\frac{1}{\sqrt{2^{N+1}}}\sum_{S'}|S'\rangle_{A}\otimes(\bigotimes^{N}_{i=0}|(-1)^{s'_{i}}\alpha\rangle)_{B},
\end{eqnarray}
is prepared.\\
\\ 
The total entangled state $|\Phi_{DPS}\rangle_{AB}$ can also be written as 
\begin{eqnarray}
|\Phi_{DPS}\rangle_{AB}&=&\bigotimes^N_{i=0}\left(\frac{1}{\sqrt{2}}\sum_{s'_{i}=0,1}|s'_{i}\rangle_{A}\otimes|\psi(s'_{i})\rangle_{B}\right)\nonumber\\
&=&\bigotimes^N_{i=0}\left(\frac{1}{\sqrt{2}}\sum_{s'_{i}=0,1}|s'_{i}\rangle_{A}\otimes|(-1)^{s'_{i}}\alpha\rangle_{B}\right)\nonumber\\
&=&\bigotimes^N_{i=0}\left(\frac{1}{\sqrt{2}}\{|0\rangle_{A}\otimes|\alpha\rangle_B+|1\rangle_{A}\otimes|-\alpha\rangle_B\}\right)^i,
\end{eqnarray}
where the states $|0\rangle$ and $|1\rangle$ form the arbitrary orthogonal basis in which Alice measures, and correspond to $s'_{i}=0$ or 1 respectively. Again, there is a one-to-one correspondence between each $s'_{i}$ and each state $|\psi(s'_{i})\rangle$ encoding that symbol, although the potential key bits $s_{i}$ are not independent of each other.\\
\\
Consequences of this non-independence are that Alice must keep track of the time intervals to which her measurement outcomes correspond, thus incrementally building her knowledge of the string $S$ via the relation $s_{i}=s'_{i-1}+s'_{i}$. Subsequent to Alice's projections in the arbitrary orthogonal basis $\{|0\rangle,|1\rangle\}$ performed on her subsystem, the resulting string of attenuated coherent pulses that form Bob's subsystem must be separated by the same time delay associated with the delay in Bob's interferometer, $\Delta t$. Bob learns a fraction $|\alpha|^{2}$ of the $s_{i}$'s in string $S$ by interfering consecutive pulses to learn their relative phases. He performs the same total measurement as in the P$\&$M description.
\section{Bob's Measurement}
Bob's measurement can be described conveniently in the framework of
generalized measurements, where the measurement statistics are given by
a positive-operator valued measure (POVM) \cite{busch94}. The POVM associates with
each measurement result $j$, a positive operator $E_j$, termed an effect or POVM element. The expectation value of the effect $E_j$ determines the probability to obtain
result $j$:
\begin{equation}
  p_j= \langle \psi| E_j|\psi\rangle\,,
\end{equation}
where $|\psi\rangle$ is the state of the system considered just before the measurement
is carried out. The effects satisfy $\sum_j E_j = I$, where $I$ is the identity operator, in order to
guarantee that the probabilities sum up to unity. In this formalism,
common projection measurements of quantum mechanical observables
are described by effects which are mutually commuting projectors onto the eigenspaces
corresponding to the measurement results. In general, e.g.\ for
indirect projection measurements, the effects are neither projectors
nor do they commute.\\
\\    
A necessary requirement for the shared bit string to be secret is that
the performed measurements must be able to detect entanglement in the
state effectively distributed between Alice and Bob in the EB scheme
\cite{curty04}. This is not possible if Bob's measurement results
correspond only to mutually commuting effects (see Appendix). This
condition applies also to the P$\&$M version of any QKD protocol,
where the measurements must be the same as in the equivalent EB
transcription. In an intercept-and-resend attack on a P$\&$M protocol where Bob's measurement is described by only commuting effects, an eavesdropper
could measure an observable which
commutes with all of Bob's effects without changing the statistics of
Bob's measurement and thus remain undetected. Therefore, a precondition for security is that some of the effects constituting Bob's measurement must be non-commuting.\\
\\
In DPSQKD Bob's measurement is associated with a total number $2^{2N}$ of
possible results and as many corresponding effects, since in the time intervals $i\in\{1, ..., N\}$, he projects onto either vacuum (no click) or a one-or-more photon state (click) in each of two detectors, $D_0$ and $D_1$. He obtains an average of $|\alpha|^{2}N$ detection events which contribute to the key. The effects constituting Bob's measurement are written as follows:
\begin{eqnarray}
G_{1}&=&|0\rangle_{4}^{1}\langle0|\otimes|0\rangle_{5}^{1}\langle0|\otimes|0\rangle_{4}^{2}\langle0|\otimes|0\rangle_{5}^{2}\langle0| ... \otimes|0\rangle_{4}^{N}\langle0|\otimes|0\rangle_{5}^{N}\langle0|,\nonumber\\
G_{2}&=&\sum^{\infty}_{n=1}|n\rangle_{4}^{1}\langle n|\otimes|0\rangle_{5}^{1}\langle0|\otimes|0\rangle_{4}^{2}\langle0|\otimes|0\rangle_{5}^{2}\langle0| ... \otimes|0\rangle_{4}^{N}\langle0|\otimes |0\rangle_{5}^{N}\langle0|,\nonumber\\
G_{3}&=&|0\rangle_{4}^{1}\langle0|\otimes|0\rangle_{5}^{1}\langle0|\otimes|0\rangle_{4}^{2}\langle0|\otimes\sum^{\infty}_{n=1}|n\rangle_{5}^{2}\langle n| ... \otimes|0\rangle_{4}^{N}\langle0|\otimes |0\rangle_{4}^{N}\langle0|,\nonumber\\
\vdots&&\nonumber\\
G_{2^{2N}}&=&\sum^{\infty}_{n=1}|n\rangle_{4}^{1}\langle n|\otimes\sum^{\infty}_{n=1}|n\rangle_{5}^{1}\langle n| ...
\sum^{\infty}_{n=1}|n\rangle_{4}^{N}\langle n|\otimes\sum^{\infty}_{n=1}|n\rangle_{5}^{N}\langle n|.
\end{eqnarray}  
Bob's measurement is thus seen to be a degenerate projection
measurement of photon number, where $[G_{i},G_{j}]=0$ for all $i,j$,
since the inner product of vacuum with a one-or-more photon state is
always zero. It is easy to show that this result does not change if
Bob's interferometer is considered as part of his measurement
apparatus. Since the action of an interferometer is unitary, the
result for the transformed effects remains the same:
$[U^{\dagger}G_{i}U,U^{\dagger}G_{j}U]=0$.  At this point the
protocol appears to be insecure! In the remainder of this section it
will be shown that the necessary condition on Bob's measurement- the non-commutativity of effects- nevertheless is met.\\
\\
For this purpose, recall that Bob's interferometer
has two input paths (path $0$ and path $1$, see Fig. \ref{figure1}). Only path $0$ is
populated, it carries the light sent by Alice in state
$|\psi\rangle_0:=|\Psi(S')\rangle_{DPS}$ (see Eq. (\ref{statedps})), while path $1$ contains the vacuum state $|0\rangle_0$ at all times. When including the
interferometer in Bob's measurement, the probability to obtain any result $j\in \{1,\,2 ..., 2^{2N}\}$ can be expressed by means of
the state $|\Xi\rangle:=|\psi\rangle_0\otimes|0\rangle_1$ of the light
entering the interferometer as:
\begin{equation}
  p_j=\langle\Xi|U^\dagger G_j U|\Xi\rangle={}_0\langle \psi |E_j
  |\psi\rangle_0 \label{pj}
\end{equation}
with
\begin{equation}
E_j:={}_1\langle 0|U^\dagger G_j U|0\rangle_1\,.\nonumber
\end{equation}
While the action of the interferometer is represented by the operator
$U$ which maps the incoming state in paths $0$ and $1$ to the the
outgoing state in paths $4$ and $5$, the new effects $E_j$ are
operators that act only on states in path $0$. The expectation
value with respect to the vacuum state in path $1$ reduces the action
of the operator $U^\dagger G_j U$ to the subspace of states in path
$0$, similarly to a partial trace. According to Eq. (\ref{pj}), the
probability for any of Bob's measurement results can thus be expressed
only in terms of the state sent by Alice using effects $E_j$. It is
well known that such a reduction of a projection-valued measure (PVM) as
given by the effects $U^\dagger G_j U$ can result in a POVM with
non-commuting effects. In fact, any POVM can be represented as a
projection of a PVM acting on a higher dimensional Hilbert space
(see the Theorem of Neumark \cite{peres90}).\\        
\\
Indeed, the resulting effects $E_j$ are not all mutually commuting and
therefore satisfy the necessary condition for security. Consider, for example, the effects  $E_2$ and $E_3$ that correspond to a click in $D_0$ in time interval $1$ and a click in $D_1$ in time interval $2$, respectively:
\begin{equation}
E_2={}_1\langle 0|U^\dagger G_2 U|0\rangle_1=\sum_{n=1}^{\infty}\frac{1}{4^{n}n!}(\hat{a}^{\dagger 0}_{0}
+\hat{a}^{\dagger 1}_0)^n|0\rangle\langle0|(\hat{a}^{0}_{0}+\hat{a}^{1}_0)^n,
\end{equation}
\begin{equation}
E_3={}_1\langle 0|U^\dagger G_3 U|0\rangle_1=\sum_{m=1}^{\infty}\frac{1}{4^{m}m!}(\hat{a}^{\dagger 1}_{0}
-\hat{a}^{\dagger 2}_0)^m|0\rangle\langle0|(\hat{a}^{1}_{0}-\hat{a}^{2}_0)^m.
\end{equation}
The commutator $[E_2,E_3]$ is given by:
\begin{eqnarray}
[E_2,E_3]&=&\sum_{n=1}^{\infty}\frac{1}{4^{n}n!}(\hat{a}^{\dagger 0}_{0}
+\hat{a}^{\dagger 1}_0)^n|0\rangle T\sum_{m=1}^{\infty}\frac{1}{4^{m}m!}\langle0|(\hat{a}^{1}_{0}-\hat{a}^{2}_0)^m\nonumber\\
&&-\sum_{m=1}^{\infty}\frac{1}{4^{m}m!}(\hat{a}^{\dagger 1}_{0}
-\hat{a}^{\dagger 2}_0)^m|0\rangle T\sum_{n=1}^{\infty}\frac{1}{4^{n}n!}\langle0|(\hat{a}^{0}_{0}+\hat{a}^{1}_0)^n,
\end{eqnarray}
where the term $T$ is defined and evaluated as:
\begin{eqnarray}
T&\equiv&\langle 0|(\hat{a}^{0}_{0}+\hat{a}^{1}_0)^n
(\hat{a}^{\dagger 1}_{0}-\hat{a}^{\dagger 2}_0)^m|0\rangle\nonumber\\
&=&\sum_{k=0}^n\sum_{l=0}^m{n\choose k}{m\choose l}(-1)^l\langle 0|
(\hat{a}^{0}_{0})^{n-k}(\hat{a}^{1}_0)^k(\hat{a}^{\dagger 1}_{0})^{m-l}(\hat{a}^{\dagger 2}_0)^{l}|0\rangle\nonumber\\
&=&n!,\,\\
&\mbox{with}&\,n=m=k\,\mbox{and}\,l=0.\nonumber
\end{eqnarray}
Note that $T=T^*$ is a non-zero real number.\\
\\
The commutator $[E_2,E_3]$ is then given by:
\begin{eqnarray}
[E_2,E_3]&=&\sum_{n=1}^{\infty}\sum_{l=0}^{n}\sum_{k=0}^n\frac{1}{16^nn!}{n\choose l}{n\choose k}\nonumber\\
&\times&\{(-1)^k(\hat{a}^{\dagger 0}_{0})^{n-l}(\hat{a}^{\dagger 1}_{0})^{l}|0\rangle\langle 0|(\hat{a}^1_0)^{n-k}(\hat{a}^2_0)^{k}\nonumber\\
&&-(-1)^l(\hat{a}^{\dagger 1}_{0})^{n-l}(\hat{a}^{\dagger 2}_{0})^{l}|0\rangle\langle 0|(\hat{a}^0_0)^{n-k}(\hat{a}^1_0)^{k}\}\nonumber\\
&\neq&0.
\end{eqnarray}
Since the operators $\hat{a}^{(\dagger) 0}_{0}$, $\hat{a}^{(\dagger) 1}_{0}$ and $\hat{a}^{(\dagger) 2}_{0}$ act on different Hilbert spaces, the matrix elements do not cancel. Therefore all terms in the sum are non-zero.\\
\\
It has therefore been shown that there do exist non-commuting effects in Bob's measurement in the EB translation of DPSQKD, i.e., $[E_2,E_3]\neq 0$, which is a necessary requirement for the detection of entanglement in the effectively distributed state. The protocol has thus been shown to satisfy a necessary condition for security, i.e. that Bob's measurement involves non-commuting effects. 
\section{Conclusion}
DPSQKD, an example of a distributed-phase-reference protocol, has here been described firstly as a P$\&$M scheme, and secondly translated into an EB scheme, thus fitting into the framework of description for a generic QKD protocol as outlined by Scarani et al.\ \cite{scarani08}. DPSQKD has been shown to satisfy a necessary condition for security, i.e., Bob's measurement involves non-commuting effects. The EB translation of DPSQKD formalised here, together with the proof that the necessary condition for security is met, can be considered a first step towards a potential unconditional security proof for the protocol based on entanglement distillation.
\ack{The authors wish to thank Valerio Scarani, Stefano Bettelli and Barry Sanders for interesting and valuable discussions. This work is based on research supported by the South African Research Chair Initiative of the Department of Science and Technology and National Research Foundation.} 
\section*{Appendix}
A theorem by Curty, Lewenstein and L\"utkenhaus \cite{curty04} states the following:\\
\\
\textbf{Entanglement as a precondition for secure QKD}\\
\textit{A necessary precondition for a set of POVM elements $F_{a}\otimes G_{b}$ together with the probability distribution of their occurance $P(A,B)$ to lead to a secret key via public communication is that the presence of entanglement in the effectively distributed state $|\psi\rangle_{AB}$ can be detected via an entanglement witness $W=\sum_{ab}c_{ab}F_{a}\otimes G_{b}$ with $c_{ab}$ real such that Tr$(W\sigma)\geq 0$ for all separable states and Tr$(W\rho)<0$ for at least one entangled state.}\\
\\
Suppose $W=\sum_{ab}c_{ab}F_{a}\otimes G_{b}$ is an entanglement witness with $c_{ab}$ real such that Tr$(W\sigma)\geq 0$ for all separable states and Tr$(W\rho)<0$ for at least one entangled state, and that $F_{a}$ and $G_{b}$ are Alice and Bob's POVM elements in a QKD protocol. Assume that in each time interval Alice projects onto the set of orthogonal states $|A\rangle$, and that Bob projects onto the set of orthogonal states $|B\rangle$. Then $W$ is diagonal in the basis $\{|A\rangle_{A}, |B\rangle_{B}\}$:
\begin{equation}
W=\sum_{A,B}\lambda_{AB}|A\rangle_{A}\langle A|\otimes|B\rangle_{B}\langle B|.
\end{equation}
Since $W$ is a witness and Tr$(W\sigma)\geq 0$ for all separable states $|\psi_{sep}\rangle$, it follows that for $|\psi_{sep}\rangle=|\alpha\rangle_{A}|\beta\rangle_{B}$
\begin{eqnarray}
&&\langle\psi_{sep}|W|\psi_{sep}\rangle\geq 0\nonumber\\ 
&\Rightarrow&\langle\alpha|_{A}\langle\beta|_{B}\otimes\sum_{A,B}\lambda_{AB}|A\rangle_{A}\langle A|\otimes|B\rangle_{B}\langle B|\otimes|\alpha\rangle_{A}|\beta\rangle_{B}\geq 0\nonumber\\
&\Rightarrow&\sum_{A,B}\lambda_{AB}\langle\alpha|A\rangle\langle \beta|B\rangle\langle \alpha|A\rangle\langle\beta|B\rangle\geq 0\nonumber\\
&\Rightarrow&\lambda_{\alpha\beta}\geq 0\ \ \forall\  \alpha,\beta
\end{eqnarray}
Since $W$ has diagonal representation $\sum_{i}\lambda_{i}|i\rangle\langle i|$ with $\lambda_{i}$ non-negative, $W$ is a positive operator. Therefore $\langle\psi|W|\psi\rangle\geq 0$ for all $|\psi\rangle$ including all entangled states and $W$ cannot be an entanglement witness. As a result a witness of entanglement in the effectively distributed state cannot be constructed, and the protocol cannot lead to a secret key via public communication.\\
\\
In the EB translation of a P$\&$M protocol, Alice is assumed to project onto a set of orthogonal states, therefore a necessary condition on the elements of Bob's POVM is that they should not all commute.
\section*{References}

\end{document}